\documentclass[12pt, epsfig]{article}
  \usepackage{amsthm,amsfonts}
  \usepackage{amsmath}
\newcommand{\bea}   {\begin{eqnarray}}
\newcommand{\eea}   {\end{eqnarray}}
\begin{document}
\renewcommand{\thefootnote}{\fnsymbol{footnote}}

\thispagestyle{empty}

\title{Extended Supersymmetric Quantum Mechanics \\ of Fierz and Schur Type}

\author{Zhanna Kuznetsova\thanks{{\em e-mail: zhanna.kuznetsova@ufabc.edu.br}}
~and Francesco
Toppan\thanks{{\em e-mail: toppan@cbpf.br}}
\\
\\
}
\maketitle

\centerline{$^{\ast}${\it UFABC, Rua Catequese 242, Bairro Jardim,}}{\centerline{\it\quad cep 09090-400, Santo Andr\'e (SP), Brazil.}}
{\centerline{
$^{\dag}${\it CBPF, Rua Dr. Xavier Sigaud 150, Urca,}}{\centerline {\it\quad
cep 22290-180, Rio de Janeiro (RJ), Brazil.}}}

\maketitle
\begin{abstract}
We discuss two independent constructions to introduce an ${N}$-extended Supersymmetric Quantum Mechanics.  The first one makes use of
the Fierz identities while the second one (divided into two subcases) makes use of the Schur lemma. The $N$ supercharges $Q_I$ are square roots
of a free Hamiltonian $H$ given by the tensor product of a $D$-dimensional Laplacian and a $2d$-dimensional identity matrix operator. We present the mutual relations among $N$, $D$ and $d$. The $mod~8$ Bott's periodicity of Clifford algebras is encoded, in the Fierz case, in the
Radon-Hurwitz function and, in the Schur case, in an extra independent function.
\end{abstract}
\vfill

\rightline{CBPF-NF-004/10}

\newpage
\section{Introduction}

Since its introduction in \cite{wit} the Supersymmetric Quantum Mechanics has found deep applications to mathematics, mathematical physics, nuclear physics and so on. An important line of research focused on its extension to $N$ independent supercharges corresponding to the square roots of the Hamiltonian operator (see \cite{{cr},{cp},{ms},{hul},{pt},{krt},{klw},{lun}}).
The connection between Extended Supersymmetry and Clifford algebras was pointed out in \cite{{cr},{pt},{krt}}. In several works \cite{{cr},{ms},{hul},{klw},{lun}} the Fierz identities \cite{brr} were used to close the $N$-extended superalgebra; in this construction the Hamiltonian can depend on a K\"ahler (for $N=2$) or hyper-K\"ahler (for $N=4$) background.\par
In this letter we show that, besides the Fierz construction (the $F$-type supersymmetry) another extension of the Supersymmetric Quantum Mechanics 
(denoted as $S$-type supersymmetry) can be based on the Schur lemma (see \cite{oku}),  expressing
the most general algebra (real, almost complex or quaternionic) which commutes with a given set of Clifford gamma matrices . \par
We consider the free Hamiltonian given by the tensor product of a $D$-dimensional Laplacian
and a $2d$-dimensional identity matrix. We write down the mutual relations among $N$, $D$ and
$d$. The $mod~ 8$ Bott's periodicity of Clifford algebras is encoded, in the Fierz case, in the
Radon-Hurwitz $G(r)$ function (see \cite{pt}) defined, for $r=1,2,\ldots,8$, by
\begin{eqnarray}&\label{period}
\begin{array}{|c|c|c|c|c|c|c|c|c|}\hline
  r& 1 & 2 & 3 & 4 & 5 & 6 & 7 & 8 \\ \hline
  G(r) & 1 & 2 & 4 & 4 & 8 & 8 & 8 & 8\\  \hline
\end{array}&
\end{eqnarray}
In the Schur case, besides the Radon-Hurwitz function, another function encodes the Bott's periodicity. It will be denoted as ``$K(r)$" and defined, for $r=0,1,2,\ldots,7$, by
\begin{eqnarray}&\label{period2}
\begin{array}{|c|c|c|c|c|c|c|c|c|}\hline
  r&0& 1 & 2 & 3 & 4 & 5 & 6 & 7 \\ \hline
  K(r) & 1 & 2 & 4 & 4 & 4 & 2 & 1 & 1\\  \hline
\end{array}&
\end{eqnarray}
$K(r)$ gives us, see \cite{{top},{kt}}, the real ($1$), almost complex ($2$) and quaternionic
($4$) properties of the $Cl(p,q)$ Clifford algebra according to $p-q=r+2~mod~8$.
\par
The Schur case is divided into two subcases. It depends on two mutually commuting sets of gamma matrices. According to which set is picked up to express the Schur (real, almost complex, quaternionic) property we end up with either the subcase $S1$ ($D$ can be arbitrarily large, but $N$ is at most $N\leq 4$) or the subcase $S2$ ($N$ can be arbitrarily large, but $D$ is at most $\leq 3$).
\par
We work with Euclidean gamma matrices. Their vectorial indices are raised and lowered with
the identity matrix. Similarly, their spinorial indices are raised and lowered in terms of a charge conjugation matrix that we can choose to be the identity. This means that,
as a practical rule, we do not need to
bother about the position (up or down) of the indices.

\section{${N}$-Extended Supersymmetric Quantum Mechanics for a free Hamiltonian}

The free Schr\"odinger equation for a $D$-dimensional non-relativistic particle reads as
\begin{eqnarray}
i\hbar\frac{d}{dt}\Psi(t,\overrightarrow{x})&=& H\Psi(t,\overrightarrow{x}),
\end{eqnarray}
where $H$ is the free Hamiltonian, given by a tensor product of a $D$-dimensional Laplacian
$\overrightarrow{\nabla}^2=\partial_{x_1}^2+\ldots+\partial_{x_D}^2$
and a $2d$-dimensional identity matrix operator ${\bf 1}_{2d}$:
\begin{eqnarray}
H&=& -\frac{\hbar^2}{2m} \overrightarrow{\nabla}^2\otimes {\bf 1}_{2d}.
\end{eqnarray}
This free dynamical system provides a realization of the ${N}$-extended Supersymmetric Quantum Mechanics if we can construct ${N}$ independent supercharges $Q_I$, $I=0,1,\ldots, N-1$, which are the square roots of the Hamiltonian $H$ and satisfy the graded algebra (the $N$-Extended Supersymmetry Algebra)
\begin{eqnarray}\label{susyext}
\{Q_I,Q_J\}&=&\frac{1}{2}\delta_{IJ}{H},\nonumber\\
\relax [H,Q_I]&=&0,\nonumber\\
\{N_F,Q_I\}&=& 0,\nonumber\\
\relax [N_F, H]&=&0,\nonumber\\
\relax \{N_F,N_F\} &=& 2 \cdot{\bf 1}_{2d}.
\end{eqnarray}
Besides $Q_I$ and $H$, the Extended Supersymmetry Algebra admits as a generator the diagonal fermion-number operator $N_F={\bf 1}_d\oplus (-{\bf 1}_d)$.
$N_F$ possesses $d$ eigenvalues $+1$ corresponding to the even states (the ``bosons") and $d$ eigenvalues $-1$ corresponding to the odd states (the ``fermions").

\section{The $F$-type realization of supersymmetry}

The $F$-type ${N}$-Extended Supersymmetry is introduced, for the free particle, through the positions
\begin{eqnarray}
Q_0&=& \frac{\hbar}{2\sqrt{m}}(\Gamma^\mu)_{ab}\partial_\mu, \nonumber\\
Q_i&=&\frac{\hbar}{\sqrt{2 m}}(\gamma^{{i}})^{\mu\nu}(\Gamma_\mu)_{ab}\partial_\nu,
\end{eqnarray}
where ${i} =1,2,\ldots, N-1$ and $a,b=1,2,\ldots, 2d$.
\par
The construction makes use of the irreducible $D$-dimensional representation of the $Cl(0,N-1)$ Clifford algebra with generators
$\gamma^{i}$ and of the $2d$-dimensional Weyl-type representation of the $Cl(0,D)$ Euclidean Clifford algebra
generated by $\Gamma_\mu$. We have\footnote{The closure of the  (\ref{susyext}) algebra
can be easily proven by making use of several Fierz identities. As an example, the vanishing of the anticommutator between $Q_i$ and $Q_j$, for $i\neq j$, requires, in particular, the vanishing of the term
$[\Gamma_\mu,\Gamma_\rho]\partial_\nu\partial_\sigma B^{ij}_{\mu\rho\nu\sigma}$,
where $
B^{ij}_{\mu\rho\nu\sigma}=
\gamma^i_{\mu\nu}\gamma^j_{\rho\sigma}+\gamma^j_{\mu\nu}\gamma^i_{\rho\sigma}
$. The term above vanishes due to the fact that $(B^{ij}_{\mu\rho\nu\sigma})^- =
B^{ij}_{\mu\rho\nu\sigma}-B^{ij}_{\rho\mu\nu\sigma}$ is antisymmetric in the exchange
$\nu\leftrightarrow \sigma$. Indeed, $(B^{ij}_{\mu\rho\nu\sigma})^-=-(B^{ij}_{\mu\rho\sigma\nu})^-$.}
\begin{eqnarray}
\{\gamma^{i},\gamma^{j} \} &=&-2\delta^{{i}{j}}{\bf 1},\quad\quad {i},{j}=1,\ldots, N-1,\nonumber\\
\{\Gamma^{\mu},\Gamma^{\nu} \} &=&-2\delta^{{\mu}{\nu}}{\bf 1} \quad\quad {\mu},{\nu}=1,\ldots, D.
\end{eqnarray}
Given $D$, the minimal value $d$, corresponding to the irreducible representation of the Clifford algebra, is unambiguously fixed and the maximal value $N_{max}$ of the supersymmetric extension can be computed.
Expressing, for $D\geq 1$,
\begin{eqnarray}
D&=& 8k+r,
\end{eqnarray}
with $r=0,1,2,\ldots, 7$,  we have,
for the minimal $d$,
\begin{eqnarray}
d &=& \frac{1}{2} G(r+1)\cdot 16^k
\end{eqnarray}
and, for $N_{max}$,
\begin{eqnarray}
N_{max}&=& 8k+G(r+1).
\end{eqnarray}
One should note the appearance of the Radon-Hurwitz function.\par
We present, up to $D\leq 16$, the table
 \begin{eqnarray}&
 \begin{array}{|c|c|c|}\hline
D  & d&N_{max}       \\ \hline
  1&1&2\\ \hline
  2&2&4\\ \hline
  3&2&4\\ \hline
  4&4&8\\ \hline
  5&4&8\\ \hline
  6&4&8\\ \hline
  7&4&8\\ \hline
  8&8&9\\ \hline
  9&16&10\\ \hline
  10&32&12\\ \hline
  11&32&12\\ \hline
  12&64&16\\ \hline
  13&64&16\\ \hline
  14&64&16\\ \hline
  15&64&16\\ \hline
16&128&17 \\\hline
\end{array}&
 \end{eqnarray}

\section{The $S$-type realization of supersymmetry}

The $S$-type ${N}$-Extended Supersymmetry for the free particle is introduced through the generators
$(\gamma^{i})_{\alpha\beta}, (\Gamma^{m})_{\alpha\beta}$ ($\alpha,\beta=1,\ldots, d$),
which satisfy the set of equations
\begin{eqnarray}
\{\gamma^{i},\gamma^{j} \} &=&-2\delta^{{i}{j}}{\bf 1},\quad\quad {i},{j}=1,\ldots, N-1,\nonumber\\
\{\Gamma^m,\Gamma^n \} &=&-2\delta^{mn}{\bf 1},\quad\quad {{m}},{n}=1,\ldots, D,\nonumber\\
\relax [\gamma^{i} ,\Gamma^{m}]&=& 0.
\end{eqnarray}
We have
\begin{eqnarray}
Q_0&=&\frac{\hbar}{\sqrt{2}}\left(
                                \begin{array}{cc}
                                  0 & \frac{1}{m}\partial_{n}\Gamma^{n} \\
                                  \partial_{n}\Gamma^{n} & 0 \\
                                \end{array}
                              \right)
,\nonumber\\
Q_{i}&=& \frac{\hbar}{\sqrt{2}}\left(
                                \begin{array}{cc}
                                  0 & \frac{1}{m}\gamma_{i}\partial_{n}\Gamma^{n} \\
                                  -\gamma_{i}\partial_{n}\Gamma^{n} & 0 \\
                                \end{array}
                              \right).
\end{eqnarray}
In the subcase $S1$ at first we fix $D$ and compute the maximal number of $\gamma^i$ matrices admitted by the Schur lemma. Conversely, in the subcase $S2$ we fix at first $N$ and
compute the maximal number of $\Gamma^n$ matrices admitted by the Schur lemma. The subcases
$S1$ and $S2$ coincide for $D=d=1$, $N=2$ and for $D=3$, $d=4$, $N=4$.
\subsection{The $S1$-subcase}

Let us fix $D\geq 1$ given by
\begin{eqnarray}
D&=& 8k+r,
\end{eqnarray}
with $r=0,1,2,\ldots, 7$.
The minimal $d$ corresponding to the irreducible representation of the Clifford algebra is
\begin{eqnarray}
d &=& \frac{1}{2} G(r+1)\cdot 16^k.
\end{eqnarray}
The maximal value $N_{max}$ of the extended supersymmetry is given by the $K(r)$ function:
\begin{eqnarray}
N_{max}&=& K(r).
\end{eqnarray}
Up to $D\leq 16$ we have the table
\begin{eqnarray}&
 \begin{array}{|c|c|c|}\hline
D  & d&N_{max}       \\ \hline
  1&1&2\\ \hline
  2&2&4\\ \hline
  3&2&4\\ \hline
  4&4&4\\ \hline
  5&4&2\\ \hline
  6&4&1\\ \hline
  7&4&1\\ \hline
  8&8&1\\ \hline
  9&16&2\\ \hline
  10&32&4\\ \hline
  11&32&4\\ \hline
  12&64&4\\ \hline
  13&64&2\\ \hline
  14&64&1\\ \hline
  15&64&1\\ \hline
16&128&1 \\\hline
\end{array}&
 \end{eqnarray}
\subsection{The $S2$-subcase}
Let us fix the number of extended supersymmetries being given by
\begin{eqnarray}
N&=& 8k+r+1,
\end{eqnarray}
with $r=0,1,2,\ldots, 7$.
The minimal $d$ is expressed by
\begin{eqnarray}
d &=& \frac{1}{2} G(r+1)\cdot 16^k,
\end{eqnarray}
while the maximal value $D_{max}$ is obtained through
\begin{eqnarray}
D_{max}&=& K(r)-1.
\end{eqnarray}
Up to $N\leq 14$ we have the table (the missing cases correspond to a Hamiltonian with no Laplacian, namely $D=0$):
\begin{eqnarray}&
 \begin{array}{|c|c|c|}\hline
N  & d&D_{max}       \\ \hline
  2&1&1\\ \hline
  3&2&3\\ \hline
  4&2&3\\ \hline
  5&4&3\\ \hline
  6&4&1\\ \hline
  10&16&1\\ \hline
  11&32&3\\ \hline
  12&32&3\\ \hline
  13&64&3\\ \hline
  14&64&1\\ \hline
\end{array}&
 \end{eqnarray}
\section{Conclusions}

In this paper we investigated two independent (for a generic value $N$) types of $N$-Extended Supersymmetric Quantum Mechanics associated with a free Hamiltonian (given by the tensor product of a $D$-dimensional Laplacian and a $2d$-dimensional identity matrix operator) and based on Clifford algebras. We investigated the mutual relations among $N$, $D$ and $d$ and proved that, in the construction based on the Fierz identities ($F$-type supersymmetry), the
Bott's periodicity of the Clifford algebra is encoded in the Radon-Hurwitz function.  For the construction based on the Schur lemma ($S$-type supersymmetry) the Bott's periodicity is also encoded in the extra function (\ref{period2}).
\par
The $S$-type supersymmetry is split into two subcases ($S1$ and $S2$). For a given $N>4$,
$d$ is the same for both $F$-type and $S2$-type supersymmetry. On the other hand, in the Fierz case the $D$-dimensional Laplacian admits $D\geq 4$ while, in the $S2$-case, the Laplacian admits at most $D=3$.\par
It is outside the scope of this letter and left for future investigations to analyze the consistency conditions for $S$-type supersymmetry for a generic Hamiltonian. In the Fierz case the constraints require a non-trivial K\"ahler ($N=2$) or hyper-K\"ahler ($N=4$) background.
\\{}~
\par {\large{\bf Acknowledgments}}
{}~\par
This work was supported by Edital Universal CNPq Proc. 472903/2008-0.

\end{document}